# Hadron-quark phase transition in the SU(3) local Nambu – Jona-Lasinio (NJL) model with vector interaction

Grigor Alaverdyan [1]

[1] Yerevan State University; galaverdyan@ysu.am

**Abstract:** *We study the hadron-quark hybrid equation of state (EOS) of compact-star matter. The Nambu—Jona-Lasinio (NJL) local SU(3) model with vector-type interaction is used to describe the quark matter phase, while the relativistic mean field (RMF) theory with scalar-isovector $\delta$ -meson effective field adopted to describe the hadronic matter phase. It is shown that the larger the vector coupling constant $G_V$ , the lower the threshold density for the appearance of strange quarks. For a sufficiently small value of the vector coupling constant, the functions of the mass dependence on the baryonic chemical potential have regions of ambiguity which leads to a phase transition in non-strange quark matter with an abrupt change in the baryon number density. We show that within the framework of the NJL model, the hypothesis on the absolute stability of strange quark matter is not realized. In order to describe the phase transition from hadronic matter to quark matter, the Maxwell's construction is applied. It is shown that the greater the vector coupling, the greater the stiffness of the EOS for quark matter and the phase transition pressure. Our results indicate that the infinitesimal core of the quark phase, formed in the center of the neutron star, is stable.*

## 1. Introduction

The study of the thermodynamic properties and constituent composition of a strongly interacting matter at extremely high density / temperature is important for understanding many phenomena both in microscopic macroscopic worlds. Many regularities in the region of low densities and high temperatures can be obtained from experiments of heavy-ion collisions. In the region of extremely high densities and low temperatures, compact stars are the main source that enriches our understanding about the structure of the matter [1-4]. It is known that the macroscopic properties of compact stars have a functional dependence on the equation of state of matter in a fairly wide range of densities, starting from the density of ordinary matter of atomic-molecular composition to values tens of times higher than the nuclear density at which various exotic structures can appear, such as pion condensate, kaon condensate, deconfined quark phase, color superconducting phase, etc.

Theoretical studies of the integral parameters of the static configurations of stars and a comparison with the results of observations make it possible to refine the equation of state of superdense matter and exclude those that contradict to observational data obtained for most of the existing models of the equation of state. The study of dynamic phenomena such as cooling of neutron stars, possible catastrophic jump-like change in the integral parameters, and the energy release associated with the latter, allows us to clarify the ideas about phase transitions and the formation of various exotic phases in the bowels of compact stars.

Over the past several decades, numerous works have appeared devoted to the theoretical study of the thermodynamic properties of deconfined quark matter and the elucidation of the possibility of its presence in the central part of compact stars. Research in this area became more intense after the discovery of the pulsars PSR J1614-2230 [5] and PSR J0348 + 0432 [6], which have a mass of about two solar masses. The existence of neutron stars with such masses indicates the stiffness of the equation of state for compact-star matter. The absence of a unified rigorous theory that would allow the equation of state (EoS) of a strongly interacting matter to be calculated in this wide range of density values leads to the need of using different models for describing the properties of matter in both hadronic and deconfined quark phases.

In many works, the authors have combined the phenomenological MIT bag model [7] for quark matter with various models for hadronic matter to obtain a hybrid EoS and, on its basis, have investigated the properties of quark-hadron hybrid stars (see, e.g., [8-19]). The MIT quark bag model was also used to study the properties of so-called strange stars [20-27], the existence of which is associated with the Bodmer-Witten-Terazawa-Itoh hypothesis of the absolute stability of strange quark matter in relation to ordinary nuclear matter [28-31].

Recently, the Nambu - Jona-Lasinio (NJL) model [32-33] was used very often to describe the quark matter. The NJL model, which was firstly proposed to explain the origin of the nucleon mass by considering spontaneous chiral symmetry breaking, was later reformulated to describe quark matter. This model successfully reproduces many features of quantum chromodynamics (QCD) [34-38]. By combining various modifications of the NJL quark model with different models for describing hadronic matter, the authors of a number of works have constructed a hybrid equation of state and, on their basis, have investigated the properties of neutron stars containing deconfined quark matter (see, e.g., [39-47]).

It was shown in Ref. [48] that, within the framework of the NJL model, the vector interaction has a repulsive character and leads to a more stiff equation of state for quark matter. The results of Refs. [15, 16, 49] have shown that in the relativistic mean field (RMF) model, taking into account the $\delta$ - meson effective field increases the stiffness of EOS for hadronic matter.

The aim of this work is to obtain a hybrid EOS of strongly interacting matter within the framework of a model in which the quark component is considered in the local SU (3) NJL model including the vector interaction term, meanwhile hadronic matter is described in the framework of the RMF model, taking into account the contribution of the scalar-isovector $\delta$ -meson field. Due to the uncertainties in the value of the vector coupling constant $G_V$ of quarks, different values of this constant are considered and the changes in the parameters of the first-order phase transition, caused by the different vector interaction strengths, are clarified.

The paper is organized as follows. In Section 2 we describe the SU(3) local NJL model with scalar and vector couplings for deconfined quark matter, and present the computational results for thermodynamic quantities of cold three-flavor quark matter in $\beta$ -equilibrium. In Section 3 we describe the RMF model with $\delta$ -meson effective field for hadronic matter in $\beta$ -equilibrium. In Section 4 we study the constant pressure phase transition (Maxwell scenario) from hadronic matter to quark matter, and construct the EOS of compact star matter for different values of vector coupling constant. Finally, in Section 5 we summarize and discuss our findings from this work.

**2. Quark Matter Phase**

*2.1. Local SU(3) NJL Model with a Vector Interaction Term*

In order to describe the deconfined quark matter, we use the local three-flavor Nambu - Jona-Lasinio (NJL) model. The Lagrangian density within the local SU(3) NJL model, including the terms of both scalar and vector interactions, has the following form:

$$\mathcal{L}_{NJL} = \bar{\psi}\left(i\gamma^{\mu}\partial_{\mu} - \hat{m}_0\right)\psi + G_S \sum_{a=0}^{8}\left[\left(\bar{\psi}\lambda_a\psi\right)^2 + \left(\bar{\psi}i\gamma_5\lambda_a\psi\right)^2\right]$$
$$- K\left\{\det_f\left(\bar{\psi}(1+\gamma_5)\psi\right) + \det_f\left(\bar{\psi}(1-\gamma_5)\psi\right)\right\} - G_V \sum_{a=0}^{8}\left[\left(\bar{\psi}\gamma_{\mu}\lambda_a\psi\right)^2 + \left(\bar{\psi}i\gamma_{\mu}\gamma_5\lambda_a\psi\right)^2\right]. \quad (1)$$

Here, $\psi$ represents quark spinor fields $\psi_f^c$ with three flavors $f = u, d, s$ and three colors $c = r, g, b$. The first term is the Dirac Lagrangian density of free quark fields with the current quark mass matrix $\hat{m}_0 = diag(m_{0u}, m_{0d}, m_{0s})$. The second term corresponds to the chiral-symmetric four-quark interaction with the coupling constant $G_S$, where $\lambda_a$ $(a = 1, 2, \ldots 8)$ are Gell-Mann matrices and $SU(3)$ group generators in the flavor space, and $\lambda_0 = \sqrt{\frac{2}{3}}\hat{I}$ ($\hat{I}$ is the identity $3 \times 3$ matrix). The third term corresponds to the six-quark Kobayashi - Maskawa - 't Hooft [50] interaction with the coupling constant $K$, which leads to the axial $U_A(1)$ symmetry breaking. This interaction is important for obtaining the mass splitting between pseudoscalar isosinglet mesons $\eta'(958)$ and $\eta(547)$. It is thanks to this term in the chiral limit ($m_{0u} = m_{0d} = m_{0s} = 0$) that the $\eta'$ meson mass grows to a finite value, while other pseudoscalar mesons, including $\eta$, remain massless. Inclusion of this interaction term in the Lagrangian makes it possible to reproduce the mass values of the $\eta$ and $\eta'$ mesons within the framework of the NJL model. The fourth term represents the vector and axial-vector channels of interaction with a positive coupling constant $G_V$.

Using the mean field approximation from the Lagrangian (1), we can obtain an expression for the functional part of the density of a thermodynamic grand potential $\tilde{\Omega}_{NJL}(T, \{M_f\}, \{\tilde{\mu}_f\})$

$$\tilde{\Omega}_{NJL} = -\frac{3}{\pi^2}\sum_{f=u,d,s}\left\{\int_0^{\Lambda}dk\, k^2\left[E_f(k, M_f) + T\ln\left(1 + e^{-\frac{E_f(k,M_f)-\tilde{\mu}_f}{T}}\right) + T\ln\left(1 + e^{-\frac{E_f(k,M_f)+\tilde{\mu}_f}{T}}\right)\right]\right\}$$
$$+ \sum_{f=u,d,s}\left[2G_S\sigma_f(T, M_f, \tilde{\mu}_f)^2 - 2G_V n_f(T, M_f, \tilde{\mu}_f)^2\right] \quad (2)$$
$$- 4K\sigma_u(T, M_u, \tilde{\mu}_u)\sigma_d(T, M_d, \tilde{\mu}_d)\sigma_s(T, M_s, \tilde{\mu}_s),$$

where $\Lambda$ is the momentum ultraviolet cutoff parameter, the need for which arises in connection with the non-renormalizability of the NJL model, $E_f(k, M_f) = \sqrt{k^2 + M_f^2}$ is the energy of quark quasiparticle of flavor $f$, and $\tilde{\mu}_f$ is expressed through the chemical potential $\mu_f$ and number density $n_f$ of a quark of flavor f as follows

$$\tilde{\mu}_f = \mu_f - 4G_V n_f. \quad (3)$$

The quark number densities are determined by the expression

$$n_f(T, M_f, \tilde{\mu}_f) = \frac{3}{\pi^2}\int_0^{\Lambda}dk\, k^2\left[\frac{1}{1+e^{\frac{E_f(k,M_f)-\tilde{\mu}_f}{T}}} - \frac{1}{1+e^{\frac{E_f(k,M_f)+\tilde{\mu}_f}{T}}}\right]. \quad (4)$$

In Equation (2) for the thermodynamic potential, $\sigma_f(T, M_f, \tilde{\mu}_f)$ ($f = u, d, s$) denotes the quark condensates $\langle\bar{\psi}_f\psi_f\rangle$, which are defined as follows

$$\sigma_f(T, M_f, \tilde{\mu}_f) = \langle\bar{\psi}_f\psi_f\rangle = -\frac{3}{\pi^2}M_f\int_0^{\Lambda}dk\,\frac{k^2}{E_f(k, M_f)}\left[1 - \frac{1}{1+e^{\frac{E_f(k,M_f)-\tilde{\mu}_f}{T}}} - \frac{1}{1+e^{\frac{E_f(k,M_f)+\tilde{\mu}_f}{T}}}\right]. \quad (5)$$

In the mean-field (Hartree) approximation, the gap equations for the constituent quark masses have the form

$$M_u = m_{0u} - 4G_S\sigma_u + 2K\sigma_d\sigma_s ,$$
$$M_d = m_{0d} - 4G_S\sigma_d + 2K\sigma_s\sigma_u , . \qquad (6)$$
$$M_s = m_{0s} - 4G_S\sigma_s + 2K\sigma_u\sigma_d .$$

The density of the grand thermodynamic potential corresponding to the quark component is determined by the expression

$$\Omega_{NJL}(T,\{M_f\},\{\mu_f\}) = \tilde{\Omega}_{NJL}(T,\{M_f\},\{\mu_f\}) - \tilde{\Omega}_{NJL}(T=0,\{n_f=0\}) . \qquad (7)$$

Denoting the quark condensates and the masses of the constituent quarks in vacuum by $\sigma_{f0}$ and $M_{f0}$, respectively, for the thermodynamic grand potential density of the quark component one can obtain

$$\begin{aligned}\Omega_{NJL} = &-\frac{3}{\pi^2} \sum_{f=u,d,s} \left\{ \int_0^\Lambda dk\, k^2 \left[ E_f(k,M_f) + T\ln\left(1+e^{-\frac{E_f(k,M_f)-\tilde{\mu}_f}{T}}\right) + T\ln\left(1+e^{-\frac{E_f(k,M_f)+\tilde{\mu}_f}{T}}\right) \right] \right\} \\ &+ \frac{3}{\pi^2} \sum_{f=u,d,s} \int_0^\Lambda dk\, k^2 E_f(k,M_{f0}) + 2G_S(\sigma_u^2+\sigma_d^2+\sigma_s^2-\sigma_{u0}^2-\sigma_{d0}^2-\sigma_{s0}^2) \\ &- 4K(\sigma_u\sigma_d\sigma_s - \sigma_{u0}\sigma_{d0}\sigma_{s0}) - 2G_V(n_u^2+n_d^2+n_s^2) .\end{aligned} \qquad (8)$$

*2.2. Betta Equilibrated Cold Quark Matter*

For stellar matter in beta equilibrium, we can assume that the temperature $T=0$. Then, the thermodynamic grand potential of a matter consisting of deconfined quarks and electrons can be written in the following form:

$$\begin{aligned}\Omega_{QP} = &\frac{3}{\pi^2} \sum_{f=u,d,s} \left( \int_0^\Lambda dk\, k^2 \sqrt{k^2+M_{f0}^2} - \int_{(\pi^2 n_f)^{1/3}}^\Lambda dk\, k^2 \sqrt{k^2+M_f^2} \right) \\ &- \sum_{f=u,d,s} n_f \sqrt{(\pi^2 n_f)^{2/3}+M_f^2} + 2G_S\left[\sigma_u^2+\sigma_d^2+\sigma_s^2-\sigma_{u0}^2-\sigma_{d0}^2-\sigma_{s0}^2\right] \\ &- 4K\left[\sigma_u\sigma_d\sigma_s - \sigma_{u0}\sigma_{d0}\sigma_{s0}\right] - 2G_V\left(n_u^2+n_d^2+n_s^2\right) \\ &+ \frac{1}{\pi^2} \int_0^{(3\pi^2 n_e)^{1/3}} dk\, k^2\sqrt{k^2+m_e^2} - n_e\sqrt{(3\pi^2 n_e)^{2/3}+m_e^2} .\end{aligned} \qquad (9)$$

Chemical potentials of quarks are expressed in terms of constituent mass $M_f$ and number density $n_f$ of a quark of flavor $f$

$$\mu_f(n_f, M_f) = \sqrt{(\pi^2 n_f)^{2/3}+M_f^2} + 4G_V n_f, \qquad (f=u,d,s) . \qquad (10)$$

The chemical potential of electrons is a function of electron number density

$$\mu_e(n_e) = \sqrt{(3\pi^2 n_e)^{2/3}+m_e^2} . \qquad (11)$$

The quark condensates $\sigma_f$ are expressed in terms of constituent mass $M_f$ and number density $n_f$ of a quark of flavor $f$ as follows

$$\sigma_f = -\frac{3M_f}{\pi^2} \int_{(\pi^2 n_f)^{1/3}}^\Lambda dk\, \frac{k^2}{\sqrt{k^2+M_f^2}}, \qquad (f=u,d,s) . \qquad (12)$$

In that stage when the neutrinos, created in the stellar core, completely come out of the star the conditions of beta equilibrium will have the form

$$\mu_d(n_d, M_d) = \mu_u(n_u, M_u) + \mu_e(n_e), \tag{13}$$

$$\mu_s(n_s, M_s) = \mu_d(n_d, M_d). \tag{14}$$

Charge neutrality condition is given by

$$\frac{2}{3}n_u - \frac{1}{3}n_d - \frac{1}{3}n_s - n_e = 0. \tag{15}$$

For the baryon number density we have

$$n_B = \frac{1}{3}(n_u + n_d + n_s). \tag{16}$$

The system of equations (6), (13), (14), (15), (16) for a given value of $n_B$ is a closed system of seven equations, the numerical solution of which makes it possible to find the constituent quark masses $M_u, M_d, M_s$ and the particle number densities $n_u, n_d, n_s, n_e$. Then, using equations (10) - (12), one can calculate the chemical potentials $\mu_u, \mu_d, \mu_s, \mu_e$ and quark condensates $\sigma_u, \sigma_d, \sigma_s$. Knowledge of these characteristics allows one to obtain the EOS of the quark phase in a parametric form. The quark phase pressure $P_{QP}$ is determined by the expression

$$P_{QP} = -\Omega_{QP}. \tag{17}$$

The energy density is expressed in terms of the thermodynamic grand potential density $\Omega_{QP}$ as

$$\varepsilon_{QP} = \Omega_{QP} + \sum_{f=u,d,s} \mu_f n_f + \mu_e n_e. \tag{18}$$

Using equation (9) for the grand-potential density, the expression for the energy density can be represented in the form

$$\varepsilon_{QP} = \frac{3}{\pi^2} \sum_{i=u,d,s} \left( \int_0^\Lambda dk\, k^2 \sqrt{k^2 + M_{f0}^2} - \int_{(\pi^2 n_f)^{1/3}}^\Lambda dk\, k^2 \sqrt{k^2 + M_f^2} \right) + 2G_S \left( \sigma_u^2 + \sigma_d^2 + \sigma_s^2 \right)$$
$$- 2G_S \left( \sigma_{u0}^2 + \sigma_{d0}^2 + \sigma_{s0}^2 \right) - 4K \left( \sigma_u \sigma_d \sigma_s - \sigma_{u0} \sigma_{d0} \sigma_{s0} \right) + 2G_V \left( n_u^2 + n_d^2 + n_s^2 \right) \tag{19}$$
$$+ \frac{1}{\pi^2} \int_0^{(3\pi^2 n_e)^{1/3}} dk\, k^2 \sqrt{k^2 + m_e^2}.$$

### 2.3. Numerical Results

In this section, we present the numerical results obtained according to the scheme described in the previous section for electrically neutral quark matter in beta equilibrium. For the numerical calculation, we have used the parameters of the NJL model given in Ref [36], $m_u = m_d = 5.5$ MeV, $m_s = 140.7$ MeV, $\Lambda = 602.3$ MeV, $G_S = 1.835/\Lambda^2$, $K = 12.36/\Lambda^5$. The coupling constant of the vector interaction $G_V$ is a free parameter of the model. In this work, we performed calculations for several values $G_V/G_S = 0; 0.2; 0.4; 0.6$.

In Figure 1 we show the numerical results for the constituent quark masses $M_u, M_d, M_s$ and the chemical potentials of particles as a function of the baryon number density $n_B$ for several values of vector coupling $G_V$. Strange quarks can exist in the system if the condition $\mu_d(n_B) \geq M_s(n_B)$ is satisfied. The minimum value of the baryon density at which strange quarks appear, or as it is commonly called, the threshold for the appearance of $s$-quarks is determined by the equation $\mu_d(n_B^{s-thr}) = M_s(n_B^{s-thr})$. Note that this threshold density depends on the vector coupling constant $G_V$.

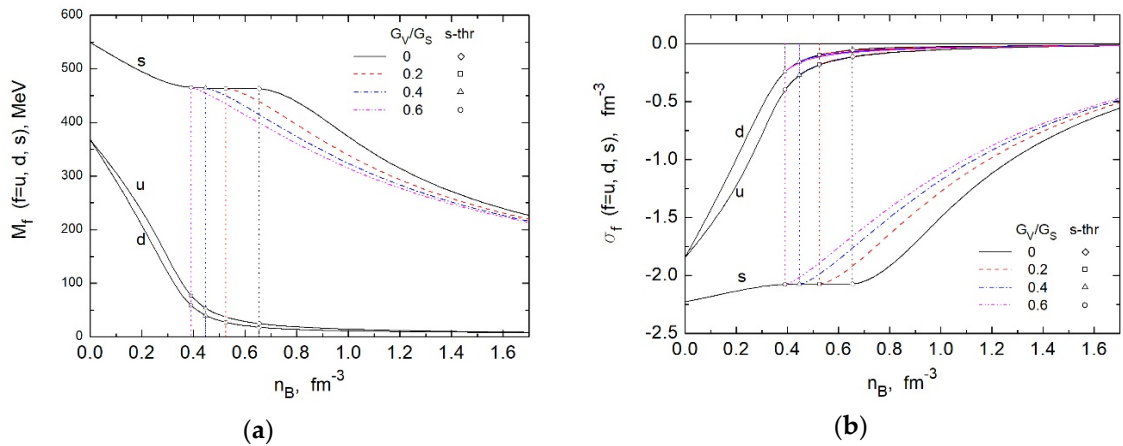

**Figure 1.** Baryon number density dependence of the constituent quark masses $M_u(n_B), M_d(n_B), M_s(n_B)$ and quark condensates in electrically neutral beta equilibrium quark matter for several values of $G_V/G_S$. In (**a**) we show constituent quark masses as a function of baryon number density. In (**b**) we show baryon number density dependence of quark condensates $\sigma_u(n_B), \sigma_d(n_B), \sigma_s(n_B)$. Vertical dotted lines show the threshold density for the appearance of strange quarks.

The parameters, such as strange-quark threshold $n_B^{s-thr}$, constituent quark masses $M_u, M_d, M_s$, baryonic chemical potential $\mu_B$, and energy per baryon $E/A$ at threshold density in electrically neutral beta equilibrium quark matter for $G_V/G_S = 0, 0.2, 0.4, 0.6$, are presented in Table 1. As can be seen from Figure 1 and Table 1, the larger the vector coupling constant $G_V$, the lower the threshold density $n_B^{s-thr}$ for the appearance of $s$-quarks. In addition, the constituent $u$ and $d$ quark masses dependent on the vector coupling very weakly. The dependences of constituent $s$-quark mass and corresponding condensate $\sigma_s$ on vector coupling are weaker in the region of densities below the threshold $n_B^{s-thr}$, and are more significant in the region above that threshold.

**Table 1.** Constituent masses, baryonic chemical potential and energy per baryon in deconfined quark matter at strange quark threshold density $n_B^{s-thr}$ for various values of vector coupling $G_V$.

| $G_V/G_S$ | $n_B^{s-thr}$ fm$^{-3}$ | $M_u$ MeV | $M_d$ MeV | $M_s$ MeV | $\mu_B$ MeV | $E/A$ MeV |
|---|---|---|---|---|---|---|
| 0 | 0.6545 | 24.87 | 17.98 | 462.94 | 1294.9 | 1142.55 |
| 0.2 | 0.5259 | 36.887 | 27.1544 | 463.241 | 1287.63 | 1156.71 |
| 0.4 | 0.4464 | 53.4407 | 40.0216 | 463.916 | 1285.07 | 1174.3 |
| 0.6 | 0.3901 | 76.7248 | 58.6001 | 465.397 | 1287.8 | 1192.72 |

The dependence of strange quarks appearance threshold density $n_B^{s-thr}$ on the value of the vector coupling constant $G_V$ is displayed in Figure 2. It can be seen that the larger the vector coupling constant, the lower the threshold for the appearance of a strange quarks in quark matter. On the ($n_B^{s-thr}, G_V$) plane of Figure 2, the region below the curve corresponds to the *ude* composition of matter, and the region above the curve corresponds to the *udse* composition.

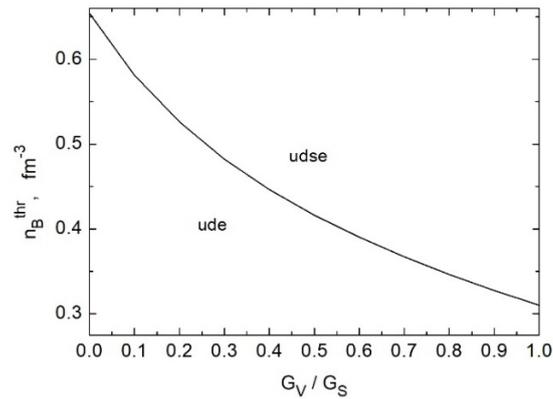

**Figure 2.** Threshold density for the appearance of strange quarks as a function of vector coupling constant $G_V$.

Figure 3 shows the results of numerical solution of the gap equations, where (**a**) and (**b**) present the constituent-quark masses $M_u, M_d, M_s$ and the pressure of the quark matter $P$, respectively, as functions of the baryon chemical potential $\mu_B$ for different values of the vector coupling constant $G_V$. As can be seen from this figure, for a sufficiently small value of the vector coupling constant $G_V$, the functions of the mass dependence on the baryonic chemical potential have regions of ambiguity. This ambiguity leads to a phase transition in non-strange quark matter with an abrupt change in the baryon number density (see the inset in Figure 3**b**). In this case, non-strange quark matter will have the instability region. With a sufficiently large value of the coupling constant $G_V$, this ambiguity disappears.

It is known that at a given value of the pressure, the greater the chemical potential, the greater the stiffness of the EOS. Figure 3b shows that the inclusion of the vector interaction term leads to an increase in the stiffness of the EOS.

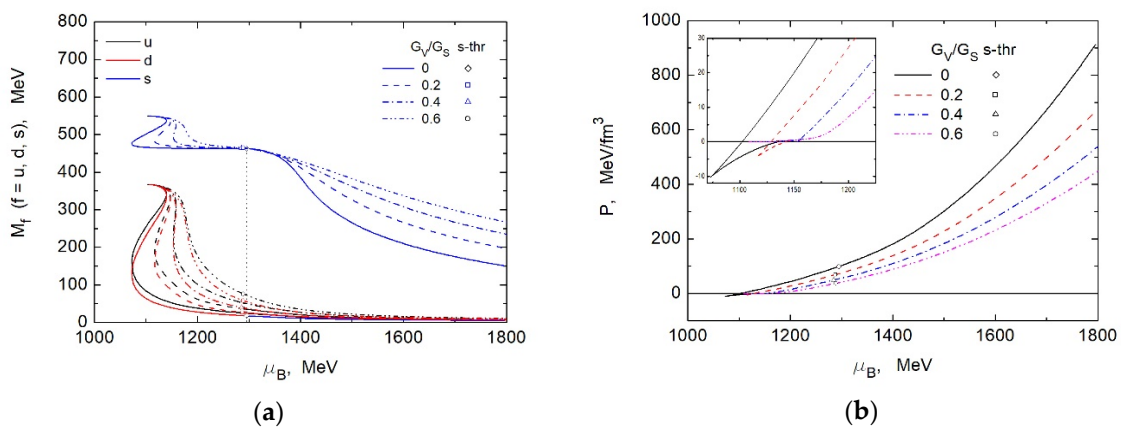

(**a**)        (**b**)

**Figure 3.** Constituent-quark masses $M_u, M_d, M_s$ (**a**) and pressure $P$ (**b**) as functions of baryon chemical potential $\mu_B$. Vertical dotted line in (**a**) denotes the strange quark threshold in case of $G_V = 0$. The inset in (**b**) shows a zoomed view of the region of the lower boundary.

In Figure 4, we show the energy per baryon $E/A = \varepsilon/n_B$ as a function of the baryon number density $n_B$ for different values of the vector coupling constant $G_V$. It is seen that the greater the vector coupling constant, the greater the energy per baryon,. Note that the value of energy per baryon in beta equlibrium quark matter exceeds the similar value in the most bound nucleus ($M_{^{56}Fe}/56$ = 930.4 MeV). This means that the hypothesis on the absolute stability of strange quark matter [28-31] within the framework of the NJL model, is not realized. In the considered version of the NJL model, the three-flavor quark matter in beta equilibrium at pressures below a certain critical value will be unstable with respect to the transition to matter with a hadronic structure. Compact stars in which the central pressure is greater than this critical value, are hybrid stars.

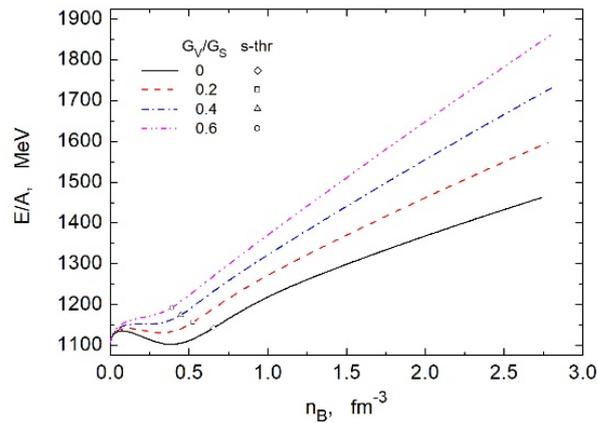

**Figure 4.** Energy per baryon as a function of the baryon number density for different values of the vector coupling constant $G_V$. Symbols, such as diamond, square, triangle and circle, denote the strange-quark formation thresholds for $G_V/G_S = 0; 0.2; 0.4; 0.6$, respectively.

It should be noted that, in the MIT bag model, at comparatively small values of the bag parameter $B$, the energy per baryon in the three-flavor quark matter at zero pressure can be less than 930.4 MeV. In this case, the three-flavor quark matter will be absolutely stable, therefore, the so-called strange stars will exist.

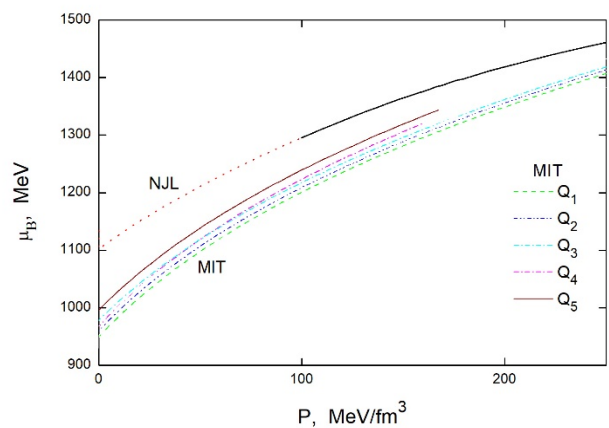

**Figure 5.** Baryon chemical potential as a function of the pressure for electrically neutral quark matter in betta-equilibrium. Comparison between the results in the NJL model at $G_V = 0$ and the MIT bag model at various values of the bag parameters, is shown in Table 2.

Figure 5 compares the results of our calculations in the framework of the NJL model without inclusion of the vector interaction term ($G_V = 0$) with the corresponding results obtained in Ref [11] in the extended MIT bag model, in which the interaction between quarks is taken into account in the one-gluon exchange approximation. The calculations in the MIT bag model were performed for five different completes of values of strange quark mass $m_s$, bag parameter $B$, and strong interaction constant $\alpha_c$, as is shown in Table 2.

Table **2.** MIT bag model parameters shown in Figure 5 for comparing with our results obtained within the NJL model

| MIT | Q$_1$ | Q$_2$ | Q$_3$ | Q$_4$ | Q$_5$ |
|---|---|---|---|---|---|
| m$_s$, MeV | 175 | 175 | 175 | 200 | 200 |
| B, MeV/fm$^3$ | 55 | 55 | 60 | 55 | 60 |
| $\alpha_c$ | 0.5 | 0.6 | 0.6 | 0.5 | 0.6 |

As we can see in Figure 5, at a given pressure, the chemical potential in the NJL model is greater than in the MIT model. The discrepancy between these values becomes even stronger when the vector interaction is taken into account. This kind of difference in the relationship between the baryon chemical potential and pressure leads to a greater value of the coexistence pressure $P_0^{NJL}$ of the two phases in the NJL model compared with the coexistence pressure $P_0^{MIT}$ in the MIT model.

## 3. Hadronic Matter Phase

Relativistic mean field (RMF) theory [51-53] is used to describe the hadronic matter phase. In the RMF approach of quantum hadrodynamics (QHD), it is believed that hadrons interact through the exchange of various mesons. In this work, we assumed that the beta equilibrium hadronic matter consists of protons (p), neutrons (n) and electrons (e), and that the exchange particles are: isoscalar-scalar meson $\sigma$; isoscalar-vector meson $\omega$; isovector-scalar meson $\delta$; isovector-vector meson $\rho$.

The relativistic nonlinear Lagrangian density for hadronic matter consisting of protons, neutrons and electrons, reads

$$\begin{aligned}\mathcal{L}_{RMF} &= \bar{\psi}_N \left[ \gamma^\mu \left( i\partial_\mu - g_\omega \omega_\mu(x) - \frac{1}{2} g_\rho \vec{\tau}_N \cdot \vec{\rho}_\mu(x) \right) - \left( m_N - g_\sigma \sigma(x) - g_\delta \vec{\tau}_N \cdot \vec{\delta}(x) \right) \right] \psi_N \\
&+ \frac{1}{2} \left( \partial_\mu \sigma(x) \partial^\mu \sigma(x) - m_\sigma^2 \sigma(x)^2 \right) - \frac{b}{3} m_N (g_\sigma \sigma(x))^3 - \frac{c}{4} (g_\sigma \sigma(x))^4 \\
&+ \frac{1}{2} m_\omega^2 \omega^\mu(x) \omega_\mu(x) - \frac{1}{4} \Omega_{\mu\nu}(x) \Omega^{\mu\nu}(x) + \frac{1}{2} \left( \partial_\mu \vec{\delta}(x) \partial^\mu \vec{\delta}(x) - m_\delta^2 \vec{\delta}(x)^2 \right) \\
&+ \frac{1}{2} m_\rho^2 \vec{\rho}^\mu(x) \vec{\rho}_\mu(x) - \frac{1}{4} \mathfrak{R}_{\mu\nu}(x) \mathfrak{R}^{\mu\nu}(x) + \bar{\psi}_e \left( i\gamma^\mu \partial_\mu - m_e \right) \psi_e,\end{aligned} \quad (20)$$

where $\psi_N = \begin{pmatrix} \psi_p \\ \psi_n \end{pmatrix}$ is isospin doublet of nucleon bispinors, $\vec{\tau}_N$ are $2 \times 2$ Pauli isospin matrices, $\sigma(x), \omega_\mu(x), \vec{\delta}(x), \vec{\rho}_\mu(x)$ are exchange meson fields at space-time point $x = x_\mu = (t, x, y, z)$ $m_N, m_e, m_\sigma, m_\omega, m_\delta, m_\rho$ are the masses of the free particles, $\Omega_{\mu\nu}(x)$ and $\mathfrak{R}_{\mu\nu}(x)$ are antisymmetric tensors of the vector fields $\omega_\mu(x)$ and $\vec{\rho}_\mu(x)$.

In the mean-field approximation, the meson quantum fields are replaced by their expectation values. The energy density and pressure of hadronic matter phase in RMF approach are writen by (For details see Ref. [54])

$$\varepsilon_{HP} = \frac{1}{\pi^2} \int_0^{k_F(n)(1-\alpha)^{1/3}} \sqrt{k^2 + (m_N - \sigma - \delta)^2} \, k^2 dk + \frac{b}{3} m_N \sigma^3 + \frac{c}{4}\sigma^4$$
$$+ \frac{1}{\pi^2} \int_0^{k_F(n)(1+\alpha)^{1/3}} \sqrt{k^2 + (m_N - \sigma + \delta)^2} \, k^2 dk + \frac{1}{2}\left(\frac{\sigma^2}{a_\sigma} + \frac{\omega^2}{a_\omega} + \frac{\delta^2}{a_\delta} + \frac{\rho^2}{a_\rho}\right) \quad (21)$$
$$+ \frac{1}{\pi^2} \int_0^{\sqrt{\mu_e^2 - m_e^2}} \sqrt{k^2 + m_e^2} \, k^2 dk \, ,$$

$$P_{HP} = \frac{1}{\pi^2} \int_0^{k_F(n)(1-\alpha)^{1/3}} \left(\sqrt{k_F(n)^2(1-\alpha)^{2/3} + (m_N - \sigma - \delta)^2} - \sqrt{k^2 + (m_N - \sigma - \delta)^2}\right) k^2 dk$$
$$+ \frac{1}{\pi^2} \int_0^{k_F(n)(1+\alpha)^{1/3}} \left(\sqrt{k_F(n)^2(1+\alpha)^{2/3} + (m_N - \sigma + \delta)^2} - \sqrt{k^2 + (m_N - \sigma + \delta)^2}\right) k^2 dk \quad (22)$$
$$- \frac{b}{3} m_N \sigma^3 - \frac{c}{4}\sigma^4 + \frac{1}{2}\left(-\frac{\sigma^2}{a_\sigma} + \frac{\omega^2}{a_\omega} - \frac{\delta^2}{a_\delta} + \frac{\rho^2}{a_\rho}\right) - \frac{1}{\pi^2} \int_0^{\sqrt{\mu_e^2 - m_e^2}} \sqrt{k^2 + m_e^2} \, k^2 dk$$
$$+ \frac{1}{3\pi^2} \mu_e (\mu_e^2 - m_e^2)^{3/2} \, ,$$

where $n$ denotes the baryon number density of the hadronic matter phase, $\alpha = (n_n - n_p)/n$ is the asymmetry parameter, $\mu_e$ is the electron chemical potential, and $k_F(n) = (3\pi^2 n/2)^{1/3}$. The renamed meson mean-fields are expressed in terms of non-vanishing expectation values of corresponding meson fields as follows:

$$\sigma \equiv g_\sigma \langle \sigma(x) \rangle, \quad \omega \equiv g_\omega \langle \omega_0(x) \rangle, \quad \delta \equiv g_\delta \langle \delta^{(3)}(x) \rangle, \quad \rho \equiv g_\rho \langle \rho_0^{(3)}(x) \rangle \, . \quad (23)$$

Parameters $a_\sigma, a_\omega, a_\delta$, and $a_\rho$ are expressed in terms of corresponding meson-nucleon coupling constant and meson mass as follows:

$$a_\sigma = (g_\sigma/m_\sigma)^2, \quad a_\omega = (g_\omega/m_\omega)^2, \quad a_\delta = (g_\delta/m_\delta)^2, \quad a_\rho = (g_\rho/m_\rho)^2 \, . \quad (24)$$

The equations for the meson mean-fields in homogeneous hadronic matter can be written in the form

$$\sigma = a_\sigma \left( n_{sp}(n,\alpha) + n_{sn}(n,\alpha) - bm_N \sigma^2 - c\sigma^3 \right), \quad (25)$$

$$\omega = a_\omega n, \quad (26)$$

$$\delta = a_\delta \left( n_{sp}(n,\alpha) - n_{sn}(n,\alpha) \right), \quad (27)$$

$$\rho = -\frac{1}{2} a_\rho n\alpha \, , \quad (28)$$

where $n_{sp}(n,\alpha)$ and $n_{sn}(n,\alpha)$ are the scalar densities of protons and neutrons, respectively, which are written in the form

$$n_{sp}(n,\alpha) = \frac{1}{\pi^2} \int_0^{k_F(n)(1-\alpha)^{1/3}} \frac{m_N - \sigma - \delta}{\sqrt{k^2 + (m_N - \sigma - \delta)^2}} k^2 dk \quad (29)$$

$$n_{sn}(n,\alpha) = \frac{1}{\pi^2} \int_0^{k_F(n)(1+\alpha)^{1/3}} \frac{m_N - \sigma + \delta}{\sqrt{k^2 + (m_N - \sigma + \delta)^2}} k^2 dk \, . \quad (30)$$

The chemical potentials of nucleons are given by

$$\mu_p(n,\alpha) = \sqrt{k_F(n)^2(1-\alpha)^{2/3} + (m_N - \sigma - \delta)^2} + \omega + \frac{1}{2}\rho,$$
$$\mu_n(n,\alpha) = \sqrt{k_F(n)^2(1+\alpha)^{2/3} + (m_N - \sigma + \delta)^2} + \omega - \frac{1}{2}\rho.$$
(31)

Taking into account the electrical neutrality condition, $n_e = n(1-\alpha)/2$, the electron chemical potential can be expressed in terms of the baryon density and the asymmetry parameter

$$\mu_e(n,\alpha) = \sqrt{\left(\frac{3}{2}\pi^2 n(1-\alpha)\right)^{2/3} + m_e^2}.$$
(32)

For neutron star matter consisting of protons, neutrons and electrons, the $\beta$ equilibrium condition without trapped neutrinos is given by

$$\mu_n = \mu_p + \mu_e.$$
(33)

In the present calculation, we use the parameter set given in Ref. [54] and listed in **Table 3**.

**Table 3.** The RMF parameter set used in this work for calculations.

| $a_\sigma$, fm² | $a_\omega$, fm² | $a_\delta$, fm² | $a_\rho$, fm² | $b$, fm$^{-1}$ | $c$ |
|---|---|---|---|---|---|
| 9.154 | 4.828 | 2.5 | 13.621 | $1.654 \cdot 10^{-2}$ | $1.319 \cdot 10^{-2}$ |

The system of equations (25), (26), (27), (28), and (33) for a given value of baryon number density $n$ is a closed set of equations, the numerical solution of which makes it possible to find the meson mean-fields $\sigma, \omega, \delta, \rho$ and asymmetry parameter $\alpha$. Knowledge of these characteristics as functions of baryon number density makes it possible to obtain a parametric form of the EOS of hadronic matter presented in (21) and (22).

**4. Compact Star Matter EOS with Hadron-Quark Phase Transition**

To construct a hybrid EOS for compact-star matter, it is necessary, in addition to the EOS of the quark phase and hadron phase separately, to have also the type of phase transition. The phase transition from hadronic matter to quark matter is principally different from the phase transition of ordinary matter consisting of atoms. In the case of ordinary atomic matter, there is only one conserved quantity, the number of atoms. In the case of a phase transition from hadronic matter to quark matter, there are two conserved quantities, the baryon number and the electric charge. Depending on the value of surface tension between hadronic matter and quark matter, one of the two scenarios of a phase transition can take place.

If the surface tension coefficient is less than a certain critical value, then the phase transition occurs with the formation of a mixed phase (Gibbs scenario). In the mixed phase, the condition of global charge neutrality is satisfied. Anyway, both the hadron phase and the quark phase are allowed to be separately charged [55]. In Gibbs (Glendenning) construction scenario, the interface between the two phases is smooth and the dependence of the energy density on pressure is a continuous function.

Otherwise, when the value of the surface tension coefficient is greater than this critical value, the phase transition is an ordinary first-order phase transition. In this case, the condition of charge neutrality is satisfied for each phase, separately. Such a phase transition occurs at a certain constant pressure value, and the characteristics of the phase transition are determined by the Maxwell construction (Maxwell scenario). The pressure dependence of the energy density is a discontinuous function with a jump at the transition pressure.

Since the value of the surface tension coefficient between quark matter and hadronic matter is still quite uncertain, it is impossible to determine which of the above two scenarios actually takes place. In this work, we assume that the surface tension between quark matter and hadronic matter is so strong that the phase transition occurs according to the Maxwell scenario.

Characteristics of the phase transition are determined according to the requirement for the simultaneous fulfillment of the conditions of mechanical and chemical equilibrium

$$P_{HP} = P_{QP} = P_0, \qquad \mu_B^{HP} = \mu_B^{QP} = \mu_{B0}, \qquad (33)$$

where $\mu_B^{HP} = \mu_n$ and $\mu_B^{QP} = \mu_u + \mu_d + \mu_s$ are the baryon chemical potentials for hadronic and quark phases, respectively.

The numerical solution of the equations (33) makes it possible to obtain the values of the transition pressure $P_0$ and the baryon chemical potential $\mu_{B0}$ at phase transition. Knowledge of these quantities will allow to calculate the remaining characteristics of the phase transition, such as: particle number densities $n_{H0}, n_{Q0}$ and the energy densities $\varepsilon_{H0}, \varepsilon_{Q0}$ for hadronic and quark phases, respectively. Graphically, these parameters can be determined by constructing a common tangent to the curves $\varepsilon_{HP}(n)$ and $\varepsilon_{QP}(n)$. The phase transition parameters can also be found by intersection of the curves $P_{HP}(\mu_B)$ and $P_{QP}(\mu_B)$ corresponding to the hadronic and quark phases.

An important parameter for a first-order phase transition, is the quantity

$$\lambda = \frac{\varepsilon_{Q0}}{\varepsilon_{H0} + P_0}. \qquad (34)$$

The value of this parameter is decisive from the point of view of stability of the infinitesimal core of the dense phase formed in the center of the neutron star [56]. If $\lambda < 3/2$, then infinitesimal quark matter core inside a neutron star is stable. The mass of the star as a function of the central pressure, at the value of the central pressure $P_c = P_0$ has a break, while maintaining a monotonically increasing character near this point. If the condition $\lambda > 3/2$ is satisfied, the mass of the star at the central pressure $P_c = P_0$ has a local maximum, which means that the infinitesimal core of quark matter newly formed in the center of the star, is unstable. During accretion of matter onto a star, when the central pressure reaches the value $P_0$, a core of finite size consisting of quark matter will be abruptly formed in the center.

**Table 4.** First order phase transition parameters for different ratios of vector and scalar coupling constants.

| $G_v/G_s$ | $P_0$ MeV/fm³ | $\varepsilon_{H0}$ MeV/fm³ | $n_{H0}$ fm⁻³ | $\varepsilon_{Q0}$ MeV/fm³ | $n_{Q0}$ fm⁻³ | $\mu_{B0}$ MeV | $\lambda$ |
|---|---|---|---|---|---|---|---|
| 0 | 150.2 | 646.88 | 0.5841 | 958.56 | 0.8128 | 1364.4 | 1.20 |
| 0.2 | 258.5 | 879.39 | 0.7449 | 1347.69 | 1.1343 | 1527.6 | 1.18 |
| 0.4 | 409.9 | 1163.47 | 0.9205 | 1515.03 | 1.3189 | 1709.4 | 0.96 |
| 0.6 | 659.5 | 1582.41 | 1.1495 | 1680.02 | 1.5420 | 1950.8 | 0.75 |

**Table 4** shows the phase transition characteristics for different values of vector coupling constant $G_V$. Note that within the framework of the model considered here, the value of the jump parameter $\lambda$ with and without inclusion of the vector interaction term, always remains less than 3/2.

In Figure 6(**a**), we show the pressure $P$ as a function of baryonic chemical potential $\mu_B$ for both the hadronic matter within the RMF model and the deconfined quark matter within the NJL model. In case of the Maxwell's construction, the phase transition is identified by the crossing points between the curves corresponding to the hadronic and quark phases in the $P - \mu_B$ plane. In Figure 6 (**b**), we plot the EOS of hybrid star matter with sharp (constant pressure) quark-hadron phase transition for different ratios of vector and scalar coupling constants: $G_V / G_S = 0, 0.2, 0.4, 0.6$. From this figure, one can find that the greater the vector coupling constant $G_V$, the greater the stiffness of the EOS for quark matter, and the greater the phase transition pressure $P_0$.

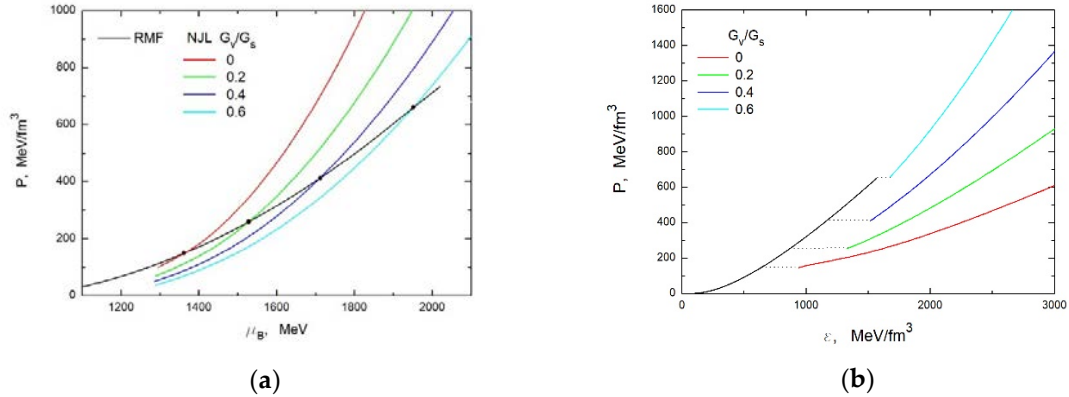

(**a**) (**b**)

**Figure 6.** (**a**) Pressure as a function of baryonic chemical potential for the hadronic matter within the RMF model and for the deconfined quark matter within the NJL model. (**b**) EOS of hybrid star matter with deconfinement phase transition for different ratios of vector and scalar coupling constants $G_V/G_S$.

## 5. Discussion and Conclusions

In this work, we have studied the phase transition from hadronic matter to quark matter, and have built the hybrid EOSs for the compact-star matter by using the Maxwell construction. In order to describe the thermodynamic properties of hadronic matter, an improved RMF theory was applied, in which, in addition to the fields of $\sigma$, $\omega$ and $\rho$ mesons, the effective field of the scalar-isovector $\delta$ - meson was also taken into account.

The thermodynamic properties of quark matter are studied in the framework of the SU(3) local NJL model, in which the vector interactions were also taken into account. For a different values of the vector coupling constant $G_V$, the gap equations for the constituent quark masses were numerically solved, and the thermodynamic characteristics of $\beta$-equilibrium charge neutral quark matter were found. The threshold values of the baryon number density, corresponding to the appearance of a strange quark, have been determined. It is shown that the greater the value of the vector coupling constant $G_V$, the lower the threshold density $n_B^{s-thr}$ of the appearance of $s$-quarks.

The influence of vector interactions on constituent quark masses and quark condensates was investigated. It is shown that the constituent masses of $u$ and $d$ quarks, and the condensates of these quarks, $\langle \bar{\psi}_u \psi_u \rangle$, $\langle \bar{\psi}_d \psi_d \rangle$ weakly depend on the value of the vector coupling constant $G_V$. The dependence of the $s$- quark constituent mass and the corresponding condensate $\langle \bar{\psi}_s \psi_s \rangle$ on the vector coupling constant $G_V$ is weaker in the region below the threshold for the appearance of the $s$- quark, and is more significant in the region above the threshold. The larger the vector coupling constant, the lower the constituent $s$- quark mass in the density region above the $s$- quark threshold.

By investigating the dependence of constituent quark masses on the baryon number density, we found that for sufficiently small values of the vector coupling constant in the region below the $s$- quark threshold, the functions $M_f(\mu_B)$ ($f = u,d,s$) have an ambiguity region. This ambiguity leads to a phase transition in non-strange quark matter with an abrupt change in the baryon number density. In this case, the non-strange quark matter will get an instability region. With a sufficiently large value of the coupling constant $G_V$, this ambiguity disappears.

We also studied the energy per baryon for $\beta$- equilibrium and charge neutral quark matter, and found that it exceeds the similar value in the most bound nucleus, $M_{^{56}Fe}/56 = 930.4$ MeV. This means that the Bodmer-Witten-Terazawa-Itoh hypothesis about the absolute stability of the strange quark matter in the framework of the NJL model, is not realized.

Comparison of the results of the NJL model with similar results of the MIT bag model showed that, at a given pressure, the baryon chemical potential in the NJL model is greater than in the MIT bag model. This kind of difference leads to a greater value of the coexistence pressure for the two phases $P_0^{NJL}$ in the NJL model compared with the coexistence pressure $P_0^{MIT}$ in the MIT bag model. It is showed that the greater the vector coupling constant, $G_V$, the greater the stiffness of the EOS for quark matter, and the greater the phase transition pressure $P_0$.

We have also calculated the density jump parameter, $\lambda = \varepsilon_{Q0}/(\varepsilon_{H0} + P_0)$, which is important from the point of view of stability of the infinitesimal core of the deconfined quark matter formed in the center of a compact star. In the model, considered in this paper, the jump parameter $\lambda < 3/2$ satisfies the stability condition for the infinitesimal core of the quark phase.

The hybrid equations of state, obtained in this work for different values of the vector coupling constant, will make it possible to calculate the integral and structural characteristics of hybrid quark-hadron stars, also to study the influence of the vector coupling constant on the properties of the star. A separate article will be devoted to these problems in the future.

**Funding**

This work was done in Research Laboratory of Superdense Star Physics at Yerevan State University and is supported by the Science Committee of the Ministry of Education, Science, Culture, and Sports of the Republic of Armenia.

**Conflicts of Interest**

The author declares no conflict of interest.